# A framework for detection and classification of events in neural activity[1]

HEMANT BOKIL†[2], BIJAN PESARAN‡, R. A. ANDERSEN‡, AND PARTHA P MITRA†

†*Cold Spring Harbor Laboratory, 1, Bungtown Road, Cold Spring Harbor, NY 11724*

‡*Division of Biology, MC 216-76, California Institute of Technology, 1200, E. California Blvd., Pasadena, CA 91125*

## Introduction

The problem of predicting behavior from observed brain activity has attracted considerable attention over the past three decades. A major impetus for this work has been the building of prosthetic devices for helping locked in patients, and indeed much work in this area uses EEG from humans to predict intent in simple motor tasks [1]. However, recent advances in chronic recording and implantation techniques[2;3] have increased the attempts to base predictions upon cortical activity recorded from implanted electrodes [4-6] instead of the EEG. Implanted electrodes have higher signal to noise ratio, and finer spatial and temporal resolution than EEG or other non-invasive techniques. While impressive progress has been achieved in predicting motor intention from recorded cortical activity [7-11], important scientific and technological issues remain.

The studies mentioned above emphasize the firing rate of multiple single units as the input to the prediction algorithm. This approach has two limitations. First, using spike rate as a summary of the signal precludes analysis of those situations where information may be present in the detailed temporal structure of spike activity. Second, acquiring and holding single cells for long periods of time remains a formidable experimental challenge. The latter issue is of special importance when clinical applications are concerned, and suggests the need for more

---

reliable control signals to drive prosthetic devices. The Local field potential (LFP) refers to the low frequency component of the recorded neural activity, which is supposed to reflect the combined synaptic activity of multiple neurons. These signals are a candidate alternative to spiking activity since they are easier to measure than single units, and have been shown to carry information about underlying motor intention [12;13]. However, algorithms to extract information from the LFP are still crude, and systematic comparisons to the amount of information available from spike trains have not yet been made.

Here, we present a general method for recognizing distinct events within neural time series data, applicable to both LFPs and spikes. The procedure, inspired by ideas drawn from speech recognition, is fully automated and can operate in real time, and does not require trial start times (in contrast to some of the related previous methods). We test its performance on a set of single unit and local field potential recordings from the lateral intraparietal area (LIP) of two macaque monkeys performing a memory-saccade task, that were first acquired and analyzed in Pesaran et al. [13]. We show that local field potentials and spikes can both be used to predict a significant proportion of the saccades to the preferred direction in advance of their occurrence, without using any information about the timing of the trials. In agreement with Pesaran et al.[13], where trial start times were specified, we find that the predictive performance of LFPs matches that of the spike rate; the current study provides detailed quantification of the relation between these two signals on a trial by trial basis. Finally, we find that the presence of single units in the recordings does not significantly affect the performance of the LFP based predictions. Our method does not require a specific relationship between the underlying neural activity and behavior, and is thus applicable with little change to many other experimental situations. Therefore it opens up the possibility of further systematic investigations regarding the information carried by spiking activity and the LFP.

**Results**

The problem we consider is that of predicting a discrete set of events, such as saccades or discrete arm movements from observed neural activity. Discrete events can be characterized by their times of occurrence, $\{t_i\}$, and labels (for example, the directions of saccades), $\{a_i\}$, and the problem is therefore that of predicting these times and labels from the observed neural activity, say, $X$. We propose a maximum likelihood solution to this problem based on estimating the log-likelihood function of the observed neural activity. We show that the log-likelihood is naturally expressed in terms of a new set of quantities which we call $2d$ cepstra. These quantities are the features underlying our analysis, and provide a parsimonious description of the neural activity. Finally, we discuss an implementation of our approach in real-time, and apply the method to the problem of predicting saccades from the LIP data mentioned in the Introduction.

*Algorithm*

While there are a number of measures that can be used to quantitate neural activity, a particularly useful measure is the spectrum, which, for a stationary process, is defined as the Fourier transform of the autocorrelation function. Neurobiological time series are not stationary. However, over short time periods of a ten-few hundred milliseconds, stationarity is still a good assumption, and a time-frequency spectrum can be defined by $S(t,f) = \left\langle \left| \tilde{X}(t,f) \right|^2 \right\rangle$, where the Fourier transform $\tilde{X}(t,f)$ is evaluated using data from a short window centered on time $t$, and the angular brackets denote a statistical average. The spectrum provides a complete characterization of second order correlations in a time series[14], both LFP and spikes, and forms the starting point of our analysis. There is however an important distinction between spikes and LFP. The LFP spectrum falls off to zero at high frequencies. In constrast the high frequency limit of the spectrum for spiking activity is the spike rate[15]. The "derivation" of the likelihood function discussed below is restricted to the case of LFP; the changes required to carry out this procedure for spiking activity are discussed towards the end of the derivation.

*Log-Likelihood*

We reasoned that in the absence of events, the observed LFP data can be characterized by a time-independent mean spectrum $\bar{S}(f)$, and assumed that the occurrence of an event at time $t_i$ causes a systematic deviation of $S(t,f)$ from this value within a window $L_i = \left[t_i - t_{Offset} - T/2, t_i - t_{Offset} + T/2\right]$. Here $t_{Offset}$ is an offset allowing for the possibility that the influence of events on the observed spectrum can either precede or follow the event in question, and $T$ is the duration of the window (Figure 1 (A)). In this case, the quantity $\hat{S}(t,f) = S(t,f)/\overline{S}(f)$ is correlated with the occurrence of events and $\left\langle \log \hat{S}(t,f) \right\rangle = 0$ for $t$ outside $L = \bigcup_i L_i$. Assuming further that distinct events are sufficiently well-separated in time that the distinct windows $L_i$ are non-overlapping (Figure 1(A)), we modeled an event of type $a_i$ occurring at time $t_i$ by a function $\Phi_{a_i}(t-t_i, f)$ which vanishes outside $L_i$. Since $\{L_i\}$ are non-overlapping,

$\Phi(t,f) = \sum_i \Phi_{a_i}(t-t_i, f)$ provides a time-frequency characterization of all events. Now, motivated by the fact that the distribution of the logarithm of the time-frequency spectrum is known to be approximately Gaussian [16], we propose the following Gaussian approximation to the log-likelihood for the residual process $\varepsilon(t,f) = \log \hat{S}(t,f) - \Phi(t,f) = \log\left[S(t,f)/\overline{S}(f)\right] - \Phi(t,f)$ (i.e. for the conditional probability density function of $\log S(t,f)$ conditioned on the events $\{t_i, a_i\}$)

$$\log P\left(\log S(t,f) | \{t_i, a_i\}\right) = -\int_{-\infty}^{\infty} df \int_{-\infty}^{\infty} df' \int_{-\infty}^{\infty} dt \int_{-\infty}^{\infty} dt' \varepsilon(t,f) K(t-t', f-f') \varepsilon(t',f'), \quad (1)$$

where $K(t-t', f-f')$ is an unknown kernel.

We now show that under certain conditions, the log-likelihood can be expressed as a sum of independent contributions from the different events. Letting $L^c$ denote the complement of $L = \bigcup_i L_i$ on the time axis, the time integrals in Equation (1) can be split up as follows:

$$\iint dt dt' = \iint_{LL} dt dt' + 2\iint_{LL^c} dt dt' + \iint_{L^c L^c} dt dt'.$$

Then, since $\Phi(t,f) = 0$ in $L^c$, Equation (1) can be written as

$$\log P(\log S(t,f)|\{t_i,a_i\}) = \iint dfdf' \left[ \iint_{LL} dtdt' \left[ 2\Phi(t,f)K(t-t',f-f')\log S(t',f') - \Phi(t,f)K(t-t',f-f')\Phi(t',f') \right] \right.$$
$$\left. + \iint_{LL^c} dtdt' \left[ \log S(t,f)K(t-t',f-f')\Phi(t',f') + \Phi(t,f)K(t-t',f-f')\log S(t',f') \right] \right], \quad (2)$$

where we have ignored contributions that are independent of the locations and labels of the events.

In the second integral in Equation (2), the regions of integration are non-overlapping ($t \in L, t' \in L^c$). Manipulation of the first integral shows it to contain two contributions, one from the same regions of integration ($t, t' \in L_i$), and another from non-overlapping regions ($t \in L_i, t' \in L_j$ for $i \neq j$). If the range of temporal correlations in the log spectrum (as captured by the kernel) is significantly smaller than $T$ (the length of the windows $L_i$), the contribution from the non-overlapping regions of integration in Equation (2) is dominated by the one from the same regions, and the log-likelihood is approximately given by

$$\log P(\log S(t,f)|\{t_i,a_i\}) = \sum_i \iint dfdf' \iint_{L_iL_i} dtdt' \left[ 2\log \hat{S}(t,f)K(t-t',f-f')\Phi_{a_i}(t'-t_i,f') \right.$$
$$\left. - \Phi_{a_i}(t-t_i,f)K(t-t',f-f')\Phi_{a_i}(t'-t_i,f') \right] (3)$$

*Feature Vectors: 2d Cepstra*

Equation (3) is a sum of independent contributions from the different events. However, it still has coupling between the two time and frequency labels which makes it difficult to use, especially because of the unknown kernel. However, by introducing the following two-dimensional Fourier transforms,

$$C_i(\phi,\tau) \equiv C(t_i;\phi,\tau) = \int_{-\infty}^{\infty} df \int_{t_i-t_{Offset}-T/2}^{t_i-t_{Offset}+T/2} dt' \log\left[\hat{S}(t',f)\right] \exp\left(2\pi i \left[\phi(t'-t_i+t_{Offset})+\tau f\right]\right),$$

$$H_{a_i}(\phi,\tau) = \int_{-\infty}^{\infty} df \int_{t_i-t_{Offset}-T/2}^{t_i-t_{Offset}+T/2} dt' \Phi_{a_i}(t'-t_i,f) \exp\left(2\pi i \left[\phi(t'-t_i+t_{Offset})+\tau f\right]\right), \quad (4)$$

$$\tilde{K}(\phi,\tau) = \int_{-\infty}^{\infty} d(f-f') \int_{-T}^{T} d(t-t') K(t-t',f-f') \exp\left(2\pi i \left[\phi(t-t')+\tau(f-f')\right]\right),$$

and using the orthogonality relations

$$\int_{-T/2}^{T/2} dt \exp(2\pi i(\phi-\phi')t) = \begin{cases} 0, & \phi \neq \phi'; \\ T, & \phi = \phi' \end{cases}$$

$$\int_{-\infty}^{\infty} df \exp(2\pi i(\tau-\tau')f) = \delta(\tau-\tau'),$$

(5)

where $\delta(\tau-\tau')$ is the Dirac delta function which is zero when $\tau \neq \tau'$ and infinite for $\tau = \tau'$, Equation (3) can be written as

$$\log P(\log S(t,f) | \{t_i, a_i\}) = \sum_i \int d\tau \sum_\phi \left[ C_i^*(\phi,\tau) \tilde{K}(\phi,\tau) H_{a_i}(\phi,\tau) + H_{a_i}^*(\phi,\tau) \tilde{K}(\phi,\tau) C_i(\phi,\tau) \right.$$
$$\left. - H_{a_i}^*(\phi,\tau) \tilde{K}(\phi,\tau) H_{a_i}(\phi,\tau) \right].$$

(6)

Here, we have ignored an overall multiplicative factor, and $C_i^*$ denotes the complex conjugate of $C_i$ (similarly for the other quantities). In Equations (4-6), the variable $\phi$ is discrete ($\phi = n/T$, where $n$ is an integer) and the variable $\tau$ is continuous. A similar analysis can be carried out for data sampled at a frequency $2F_{Nyquist}$: the only change is that the variable $\tau$ is also discrete ($\tau = m/F_{Nyquist}$, $m$ an integer) and the integral over $\tau$ in Equation (6) is replaced by a sum over $\tau$. Note that the log-likelihood on Equation (6) is not only a sum of independent contributions from the different events, it is also the sum of independent contributions from the different values of the $\phi$ and $\tau$ variables. We next show how the properties of log spectrum allow dropping the kernel from Equation (6) and lead to the final expression for the log-likelihood.

In Equation (6) the approximate log-likelihood has been expressed in terms of feature matrices $C_i$ which are two-dimensional Fourier transforms of the logarithm of the time-frequency spectrum. The transformation to these variables has the following advantage: if the time-frequency spectrum is broad-band in frequency and slowly varying in time, as is true of most neurobiological data, $C_i(\phi,\tau)$ is narrowly localized in the $\phi-\tau$ plane. This is seen in Figure 1 (C,E) which shows the absolute value of the mean $C_i(\phi,\tau)$ for saccades to the preferred and anti-preferred directions. Furthermore, it is known that estimates of the spectrum are known to be approximately uncorrelated for frequencies that differ by twice the bandwidth [16]. If, as assumed in deriving Equation (5), temporal correlations fall off rapidly over the width of the window, $\tilde{K}(\phi,\tau)$ will have a broad distribution in the $\phi-\tau$ plane. Based on these observations, we truncate the sums over $\phi$ and $\tau$ can be truncated to $\phi \leq \phi_c$ and $\tau \leq \tau_c$ (the choice of $\phi_c$ and $\tau_c$ being guided by the data under consideration), and assume that $\tilde{K}(\phi,\tau)$ is a constant over this range. Then, $\tilde{K}(\phi,\tau)$ can be absorbed into a multiplicative constant. Now, defining $\vec{\sigma}(t_i)$, and $\vec{h}(a_i)$ as complex one-dimensional vectors obtained by concatenating the components $C_i(\phi \leq \phi_c, \tau \leq \tau_c)$ and $H_{a_i}(\phi \leq \phi_c, \tau \leq \tau_c)$, the approximate log-likelihood can be written as

$$\log P(\log S(t,f) | \{t_i, a_i\}) = \sum_i \left[ 2\vec{\sigma}(t_i) \cdot \vec{h}(a_i) - |\vec{h}(a_i)|^2 \right]. \qquad (7)$$

Equation (7) is our final expression for the log-likelihood. The "maximum likelihood" solution to the event detection problem is then given by maximizing the right hand side of Equation (7) with respect to $\{t_i, a_i\}$, under the constraint that the windows $\{L_i\}$ used to compute the feature vectors are non-overlapping i.e. $|t_i - t_j| > T$ for all $i, j$. Note that if training data is available, the maximum likelihood estimate of the parameters $\vec{h}(a)$ is given by $\langle \sigma(t_i^a) \rangle$ where $t_i^a$ are the known times for events of type $a$. Thus, $\vec{h}(a)$ is the mean feature vector for events of type $a$, as is implicit in Equation (1).

The feature vectors in Equation (7) are based on the two-dimensional Fourier transform of the log spectrum. The use of such Fourier transforms is related to an approach that is common in speech recognition, where the Fourier transform of the instantaneous log spectrum viz. $C(\tau) = \int dt \log S(t, f) \exp(2\pi i \tau f)$, along with its low order derivatives is used to construct feature vectors. $C(\tau)$ is known as the cepstrum, and the variable $\tau$ is known as the quefrency [17]. In analogy with this nomenclature, we refer to the two-dimensional Fourier transforms $C_i(\phi, \tau)$ as $2d$ cepstra. While the use of the cepstrum and its derivatives can be thought of as an expansion of the time dependent cepstrum in terms of powers of time, the use of the $2d$ cepstrum can be thought of as an expansion in terms of sines and cosines. The advantage of the $2d$ cepstrum is that dynamics is automatically included in the feature vector rather than having to incorporate it via time derivatives.

*Real-time implementation*

The solution suggested above involves maximization over all possible times and labels, whereas in practice events have to be detected as data is gathered. If maximization of Equation (7) is carried out locally which marching through the data, there is a danger of locating spurious maxima and missing the true maxima. To avoid this possibility, we incorporate label-dependent thresholds, $\{\mu(a_i)\}$, on the log-likelihood, and implement the algorithm as follows:

1) $t_1$ and $a_1$ are the earliest time and the corresponding label for which:

   a) $t_1$ is a local maximum of $2\vec{\sigma}(t).\vec{h}(a_1) - |\vec{h}(a_1)|^2$,

   b) $2\vec{\sigma}(t_1).\vec{h}(b) - |\vec{h}(b)|^2 < 2\vec{\sigma}(t_1).\vec{h}(a_1) - |\vec{h}(a_1)|^2$ for all $b \neq a_1$, and

   c) $2\vec{\sigma}(t_1).\vec{h}(a_1) - |\vec{h}(a_1)|^2 > \mu(a_1)$.

2) Subsequently, $t_i$ and $a_i$ are the earliest time greater than $t_{i-1}$ and the corresponding label for which:

   a) $t_i$ is a local maximum of $2\vec{\sigma}(t).\vec{h}(a_i) - |\vec{h}(a_i)|^2$,

   b) $2\vec{\sigma}(t_i).\vec{h}(b) - |\vec{h}(b)|^2 > 2\vec{\sigma}(t_i).\vec{h}(a_i) - |\vec{h}(a_i)|^2$ for all $b \neq a_i$,

c) $2\vec{\sigma}(t_i).\vec{h}(a_i) - |\vec{h}(a_i)|^2 > \mu(a_i)$, and

d) $|t_i - t_{i-1}| > T$.

Note that if for a particular label $'a'$, $\mu(a)$ is set to be too high, no events of type $'a'$ will be predicted; on the other hand, if $\mu(a)$ is set to be too low, too many will be predicted. These parameters can therefore be adjusted to provide a desired total number of predictions for events of each type. However, since the number of recordings included in the application discussed here is large, we simply fixed the thresholds in the manner discussed in Appendix. Finally, since this is a detection problem, we define prediction to be correct if it occurs within some pre-specified window around a real event with the same label. The duration of this window is a parameter in the algorithm. This completes our derivation.

It was mentioned previously that the spike rate is the high frequency limit of the spike spectrum[15]. It is therefore necessary to separate the contribution from the spike rates and the contribution from the spectral shape of spiking activity. The latter can be quantified by the normalized quantity, $S(t,f)/R(t)$, where $R(t)$ is the spike rate. The above derivation can then be repeated to show that Equation (7) is applicable to spiking activity where the $2d$ cepstra are computed from Equation (4) with $\hat{S}(t,f) = \dfrac{S(t,f)/R(t)}{\overline{S}(f)/\overline{R}}$.

Here, $\overline{R}$ and $\overline{S}(f)$ are the mean spike rate and the mean spectrum in the absence of events. Similarly, if the spike rate $R(t)$ varies slowly over $\{L_i\}$, Equation (7) can be used for predictions based on the spike rate itself, where the feature vector taken to be the one-dimensional Fourier transform of the logarithm of $\dfrac{R(t)}{\overline{R}}$ evaluated over $\{L_i\}$. We suggest therefore that leading components of the $2d$ cepstrum for spikes be tagged on to the Fourier transform of the logarithm of the spike rate to provide a feature vector, and that the algorithm stated above be used as the basis for predictions based on spiking activity.

Finally, we mention that maximizing the log-likelihood subject to the constraints off non-overlapping windows is equivalent to minimizing the negative log-likelihood subject to these constraints. This suggests a connection between the approach adopted here and the risk minimization approach in decision theory[18]. In that approach, a risk function is constructed as a sum of the negative log-likelihood and additional terms that enforce known constraints on the events. The advantage of this approach over the more conventional Bayesian approach[19] is that constraints are less influential that prior probabilities in determining the solution. Furthermore, prior probabilities are in general unknown and difficult to estimate. The risk minimization approach is therefore particularly relevant when data is scarce.

*Application*

To test our algorithm, we attempted to predict saccades to the preferred direction from recordings of LFP and single unit activity acquired by Pesaran et al. [13] from area LIP of macaque monkeys while they performed a memory-guided saccade task. In this task, the monkeys were required to make saccades to remembered locations of targets flashed at one of eight positions arranged in a circle on a screen (see Appendix for further details on the behavioral paradigm). By analysing recordings in which single units were found, Pesaran et al. [13] observed that the firing rate and the gamma band $(30-90Hz)$ LFP power during the memory period are highest for one these directions (the preferred direction; the opposite direction is referred to as the anti-preferred direction), and found sharp directional tuning for both the gamma band LFP power and the spike rate. We found similarly sharp directional tuning for the substantially larger set of recordings under study here (see Appendix), indicating that there is little discriminability within the non-preferred directions. Since, in addition, the memory period activity in the anti-preferred direction is not very different from baseline activity, we restricted our attempts to predicting saccades in the absence of timing information to the preferred direction alone. As discussed next, that even this simplified problem is qualitatively different from the problem in which trial start times are used.

*Quantifying the performance: Need for different metrics when trial times are not supplied*

One of our initial findings is that a different set of performance metrics need to be defined for detecting and classifying events, as opposed to simply classifying trials (which is the usual case, e.g. in Pesaran et. al. 2002). In the latter case, percentages of correct or incorrect classifications provide an adequate characterization, as specified for example through a confusion matrix. In the present case, however, events need to be located in time before being classified. The events being located and classified occur comparatively infrequently in time, and do not necessarily have prominent signatures for any given event (contrast this with say spike detection and waveform classification from extracellular recordings, which is an analogous problem). For example, in the present instance, we might be locating ten events of one type that occur over a twenty minute interval.

While there are a number of ways to characterize the extra degree of freedom introduced by the event detection stage, we found it convenient to quantify the "attempt frequency" (AF). This is the average rate (events per unit time) at which the detector reported events, whether they be correct or incorrect predictions. Note that in our algorithm, the detection and classification are carried out simultaneously by means of a time dependent likelihood (or risk) function (contrast this with the problem of spike detection mentioned earlier, where detection and classification are distinct). Note that the "attempt frequency" can be adjusted by means of the thresholds on the log-likelihood. For a fixed attempt frequency, one may obtain classification and misclassification rates in the same way as one does for the more conventional case, with the caveat that events are detected within a non-zero time window.

Another quantity we introduce is the "null positive" (NP) rate, namely the fraction of events that will be correctly detected and classified simply by chance. This rate is nonzero, since the attempt frequency is nonzero. It provides the baseline for the detector: the performance of the detector may be judged by how much larger the true positive rate is compared to the null positive. Note that NP is a small number, since the events being detected occur fairly rarely: the probability that a random detector will locate the events correctly is therefore quite small. The calculation of NP as a function of AF is discussed in Appendix.

*Prediction of Saccades*

Since there is only one type of event to predict i.e. saccades to the preferred direction, the algorithm described in the previous section was applied with just one type of event (labeled P), as shown in Figure 1 (F,G). Each recording was divided into two halves of equal duration and the first half was considered the training set and the second the test set. The mean feature vector $h(P)$ and the threshold $\mu(P)$ were determined from the training set using a cross-validation procedure to reject outliers (see Appendix). The feature vectors for the LFP were based on $2d$ cepstra. The use of $2d$ cepstra to model spiking activity was precluded in these recordings by the high variability of the single-trial spike spectra [13]. This is a consequence of the limited number of spikes in each trial, and suggests that most of the behavioral information in these experiments in contained in the firing rate. As discussed at the end of the derivation, the Fourier transform of the logarithm of the spike rate can be used as a feature vector for combining the spike rate and spike cepstra. However, since the spike cepstra are not being used here, and for consistency with previous work [13], the algorithm was implemented with the rates themselves taken to be the feature vectors. The duration of the window over which the feature vectors were computed was set to be $T = 1000 ms$, and the offset $t_{Offset}$ was chosen as the mean time between the saccade and the beginning of the memory period. Since, $t_{Offset} \simeq 1100 ms$, this ensured that the windows preceded the saccades. The parameters used to compute the LFP spectra and the spike rates, and the chosen dimensions of the cepstral feature vectors are discussed in Appendix. Finally, a prediction was considered correct if it occurred within $\pm 250 ms$ of a known saccade to the preferred direction.

The results of the analysis are displayed in Figure 2, in terms of the fraction of correct predictions or true positives (TP: left y-axis), the attempt frequency (AF: right y-axis), and the null positive rate (NP: left y-axis; see Appendix for calculation). The results show that the method described here performs significantly better than the null case corresponding to a random detector with well over 50% of the saccades being correctly predicted from the LFP and approximately 50% being correctly predicted from the spiking activity. The method used here to set thresholds on the log-likelihood (see Appendix) led to varying AF and TP for the different groups. However, in each case, the method generated a reasonable number of correct predictions (TP). In addition, since NP is a computed from AF (see Appendix), the important quantity is the ratio of TP to NP. This ratio is approximately 4.5 for LFPs without single units, 6.25 for LFPs with single units and 6.5 for single unit activity, indicating that predictions from the LFP recorded at sites with single units compare favorably with those from the spike rate. The slight difference in the predictions from LFP recorded at sites without single units and those recorded at sites with single units is attributable to the heterogeneity in the quality of the recordings – note that there were over 90 recordings from sites without single units as opposed to only 28 sites at which single units were found (see Appendix). Note also that typical TP versus AF curves are likely to exhibit non-linearity so that at AF increases faster than TP. In fact, by choosing the thresholds differently for the LFP recordings we were able to get TP to NP ratios to be very similar for all three cases (results not shown). Finally, we note that the AF values in bottom panel of Figure 2 correspond to the number of false detections being approximately equal to or twice the total number of saccades.

The above conclusion of essential equivalence between the three classes was borne out from the performance on individual recordings shown in Figure 3. This figure displays the sorted values of the true positives for analyses based on (a) LFPs recorded at sites without single units, (b) LFPs recorded at sites with single units, and (c) Spike rates. While there was considerable variation in performance across recordings, the distributions for the three categories were quite similar. The attempt frequency values are not displayed: they were found to be negatively correlated with the fraction of true positives ( $P < 0.05$ ;t-test), an indication of the heterogeneity in the quality of the recordings.

Finally, the distributions of the errors in timing are displayed in Figure 4 pooled across all recordings. The distribution obtained from the LFP, and that obtained from the spike rate were found to have similar widths, approximately $70ms$ for the LFP and $85ms$ for the spikes, indicating once again that the dynamic LFP spectra and spike rate contain the similar predictive power in our data. The figure also indicates that increasing the tolerance in the timing error (set to $\pm 250ms$, see above) will not increase the quality of the predictions. There is a slight bias in the timing prediction. This is a consequence of the fact that the point process model assumed in deriving the algorithm is only approximately valid for the windows used to calculate the feature vectors in this case. In particular, the model assumes that events affect the observed time-series only within the windows used to define the feature vectors (see Appendix). In contrast, the elevation of the gamma band power in the LFP and the spike rates persist outside the windows chosen in our analysis. The bias can be removed either by increasing the duration of the windows or by including more events in the detection. The first possibility is limited by our aim of keeping a causal window between the windows used to compute the feature vectors and the event locations. We leave the second possibility for future work.

The previous analysis did not use information about the timing of the trials. To compare the results of our approach with those reported in Pesaran et al. [13], we also analyzed the problem where the timing of the saccades was known, and the task was to classify them into two categories based on the log-likelihood: (i) saccades to the preferred direction, (ii) saccades in non-preferred directions. In this case, we defined $h(O)$ as the mean feature vector for saccades to the anti-preferred direction, computed with the same parameters as before $(T = 1000ms, t_{Offset} \simeq 1100ms)$. The results are displayed in Figure 5. $"P"$ denotes the fraction of correct predictions for saccades to the preferred directions, $"A"$ denotes those for saccades to the anti-preferred direction, and $"R"$ denotes those for saccades to the remaining directions. Once again classifications based on the $2d$ cepstrum of the LFP were comparable to those based on the spike rates. More than 90% of the trials to the preferred and anti-preferred direction were classified correctly based on the LFP. When spike rates were used, the performance was slightly worse going down to 80% for the preferred direction. Furthermore, a substantial fraction of the saccades to directions other than the preferred and anti-preferred directions were classified as belonging to the anti-preferred class, confirming that there is little discriminability within trials to the non-preferred directions. These results are similar to those obtained by Pesaran et al. [13] from recordings at sites with single units. Their results for the LFP were based on using the power at a single frequency (that varied between recordings) which provided the best discrimination between trials to the preferred and anti-preferred directions. We mention in passing that we also attempted classifications based on all eight directions. The results for this case (not shown) confirmed a degradation in performance, lending further support to our approach of attempting to predict the saccades to just the preferred direction when timing information is absent. Note that the improved quality of these results compared to those displayed in Figure 2 confirms the idea that event detection and classification without using trial specific timing information is a qualitatively different, and a substantially more difficult problem than the one in which external timing information is supplied. Finally, we mention that we also attempted classifications based on all eight directions. The results for this case (not shown) confirmed a substantial degradation in performance. This lends further support to our approach of attempting to predict the saccades to just the preferred direction when timing information is absent.

Taken together, the results of our analysis indicate that the LFP spectra and the spike rates have comparable predictive power for saccade plans in LIP even when timing information is not supplied externally. The presence or absence of single units in the recordings did not correlate with the quality of the predictions. The advantage of LFP spectra over spikes, in the prosthetic context, is the ease of recording. For a particular performance level, there were always more LFP recordings without single units than recording with single units (Figure 3). For example, the ratio of TP to NP was greater than 10 in more than 30 LFP recordings but in just 10 single unit recordings.

**Discussion**

The only previous study aimed specifically at comparing predictions from LFP with those from spikes was that of Mehring et al. [12]. Using feature vectors based on the spike rate and the evoked potential, and knowledge of trial times, these authors showed that LFPs from the primary motor cortical area contain as much information about arm movements in a center-out task as that contained in the spikes. As discussed previously, prediction without use of the trial structure is a substantially more difficult problem than that where such knowledge is used.

Our work adds to the evidence provided in [12;13] of the equivalence between spikes and LFP. In addition, by providing a unified framework for detection of transient events underlying both point-like and continuous neural data our work provides a bridge between predictions based on spiking activity, and those based on continuous signals such as EEG and LFP. The method does not require trial specific timing information, nor does it require any particular model relating the observed neural data to the underlying activity. It is therefore easily applicable to other behavioral situations. We have shown that for the LIP experiments under consideration, predictions from the local field potential match those from the spikes, and that the presence or absence of isolated single units in the recordings makes little difference to the quality of the LFP based predictions. Coupled with the fact that LFPs are much easier to record than spikes, our results suggest that LFPs could supplement or replace spikes as the basis for predictions. We emphasize that while we have characterized the spiking activity in terms of the firing rate, this could be augmented with the $2d$ cepstrum. However, for the data under consideration we did not expect significant performance gains from using this procedure since most of the information about saccades is contained in the spike rate (see Appendix).

One concern may be that our classifier is contaminated by activity evoked by the visual response to target presentation rather than the memory period activity per se. We should note that the main goal of this work was to develop a suitable algorithm to detect and classify events from the LFP signal, irrespective of the precise origin of these events; therefore, this concern is not of central importance in the present work. This was also the reason that we did not attempt a systematic comparison of our choice of feature vector with simpler choices (for example, with a feature vector composed of the LFP power in distinct frequency bands), or a systematic test of the dependence on the various parameters e.g. the bandwidth, the duration of the window used to compute the feature vector. However, to assess the method in the context of the memory period per se, we repeated our calculations on a subset of the recordings choosing the windows for computations of the 2D cepstra to follow the visual transient, and found results (not shown) that were comparable to the ones stated above.

Several issues have not been addressed in this study. The time between consecutive saccades here was required to be at least $1000 ms$. However, in natural viewing a saccade occurs approximately once every 200 – 300 ms. To apply our method to such a case would require that the feature vectors be computed over windows of duration less than $200 ms$. We did not attempt to test our method in detail with such small windows, reasoning that this issue is best addressed with multi-electrode and multicellular recordings. The preferred direction of the LFP and single units recorded at a particular site was almost always identical, and these directions spanned most of the eight directions allowed in this study. It is therefore likely that the same will be true of multi-electrode recordings of LFP and that such recordings would also allow prediction of saccades to multiple directions. We have also ignored the problem of predicting other behavioral markers. This problem is perhaps best treated by combining our approach with more detailed probabilistic models of the underlying events. Finally, applying this technique to other experimental paradigms, particularly with the aim of testing the generality of the equivalence of spikes and LFPs found in our study, and implementing it in real time, remain important issues for future work.

**Appendix**

*Behavioral paradigm*

The behavioral paradigm in these experiments was as follows: at the start of the trial a spot of light appeared at the center of a screen. The monkey was required to fixate on this spot. Once fixation was acquired a peripheral target was flashed for approximately $100 ms$ at one of eight locations arranged in a circle around the fixation spot. The monkey was required to maintain fixation and hold the target location in memory until the fixation spot turned off. This was the "go" signal to the monkey to perform a saccade to the remembered location of the target. The mean time between the "go" signal and the target presentation was approximately $1000 ms$. We had available for analysis four channels of local field potential recordings from 88 sites from the first monkey, and from 44 sites in the second monkey. Spike data consisted of spike times for 16 isolated single units from 12 sites in the first monkey, and 16 from 16 sites in the second monkey [for details of spike sorting, see 13].

*Directional Tuning*

We calculated the principal components of the LFP data, and determined LFP tuning curves from the mean gamma band (25−90 Hz) power in the first principal component for a 500*ms* window immediately after the onset of the cue. A total of 113(86%) recordings were found to exhibit significant differences between the power for trials in the preferred direction and those in the anti-preferred direction ($P<0.05$; t test). For the single unit recordings, tuning curves were constructed using the mean spike rate in the same window, and 30(93%) cells were found exhibiting significant differences in the rate between trials to the preferred and anti-preferred directions ($P < 0.05$; ANOVA). The analysis of this paper involved these well-tuned recordings. Finally, 27(84%) of the recordings with single unit activity showed significant differences in gamma band LFP power between trials to the preferred and anti-preferred directions. The choice of the first principal component of the LFP is not critical to our analysis. In the recordings under study, the four LFP recordings on the tetrodes are so similar that any one of these could have been used as well. In fact, taking the average of the LFP recorded at the four tetrodes gave results that were very similar to those reported here.

*Training*

As stated in the Applications section, each recording was divided into two halves with equal number of trials, and the first half was considered the training data and the second was considered the test data. The mean time difference $t_{Offset}$ between onset of the memory period and the saccade for all the trials was calculated in each training dataset. Then, segments of data (both spikes and first principal component of the LFP) of $T = 1000ms$ duration centered at time $t_{Offset}$ before each saccade to the preferred direction in the training data were extracted. Similar segments extracted for saccades to the anti-preferred directions. The extracted segments of LFP data were used to calculate the time-frequency spectra using 300*ms* moving windows, stepped through 50*ms*.

We used the multi-taper method [16] with 5 tapers, which corresponds to a bandwidth of $\pm 10 Hz$. The mean (baseline) spectrum was computed from the training half of each recording using the same parameters. The spectra were then used to compute $2d$ cepstra using Equation 4. In a similar manner, extracted segments of spiking activity were used to compute spike rates by convolving the spike times with a Gaussian kernel (width $50ms$), stepped through $5ms$.

With these $2d$ cepstra in hand, we then computed $h(P)$ and $\mu(P)$ for each LFP recording using the following cross-validation procedure: we computed average $2d$ cepstra for saccades to the anti-preferred direction, and constructed a one-dimensional feature vector from them, say, $h_{training}(O)$. Then, we dropped one saccade to the preferred direction in turn and computed the average $2d$ cepstra from the remaining preferred directions, and the corresponding one-dimensional feature vector, say, $h_{training}(P)$. Subsequently, we tabulated all the saccades to the preferred direction satisfying $2\vec{\sigma}(t_i).\vec{h}_{training}(P) - |\vec{h}_{training}(P)|^2 > 2\vec{\sigma}(t_i).\vec{h}_{training}(O) - |\vec{h}_{training}(P)|^2$, where $t_i$ are the times of these saccades and $\vec{\sigma}(t_i)$ are the one-dimensional feature vectors constructed from the $2d$ cepstra corresponding to these dropped saccades. Saccades which do not satisfy this condition were considered outliers. The $2d$ cepstra of those that did satisfy the above condition were therefore averaged to give $h(P)$. A similar cross-validation analysis carried out for the anti-preferred directions gave the mean feature vectors $h(O)$. In computing $h(P)$ and $h(O)$ for the LFP, it was necessary to fix the dimensionality of the feature vectors. This was done by first plotting $2\vec{\sigma}(t).\vec{h}_{training}(P) - |\vec{h}_{training}(P)|^2$ in the neighborhood of the a few saccades to the preferred direction with varying choice of dimensionality. We found that the log-likelihood function changed by less than $0.0001$ when the cut-off dimensions exceeded 5 in the $\phi$ direction and 10 in the $\tau$ direction. This choice was fixed for all subsequent analyses. To complete the training process, we set

$\mu(P) = \min_i \left\{ 2\vec{\sigma}(t_i).\vec{h}_{training}(P) - |\vec{h}_{training}(P)|^2 \right\}$ where the maximum is over all the saccades tabulated earlier

as correctly classified. This is the minimum threshold required to ensure that the dropped saccades can be correctly predicted from the remaining ones by the algorithm stated in the main body of the paper. Finally, $h(P), \mu(P)$ and $h(O)$ were computed for spiking activity using the same procedure.

*Estimate of noise prediction*

Consider a detector that consists of $N$ independently placed windows of duration $\Delta T$. A correct prediction is said to occur when an event falls within one of the detector windows. $\Delta T$ can therefore be interpreted as the allowed error in the timing prediction (chosen to be $500 ms$ in the analysis described above). If the events are distributed as a Poisson process of mean rate $\lambda$, and the detector windows are placed randomly, the expected number of correct predictions by chance is $N \lambda \Delta T$. Thus, the expected *fraction* of correct predictions is $N \Delta T / T$, where $T$ is the length of the recording. $N/T$ is the number of events per unit time i.e. an attempt frequency. This formula was used to compute values of NP in Figure 2.

**Acknowledgements**: We acknowledge support from DARPA, the McKnight Foundation, Swartz Foundation, NIH (R01 MH62528-02 and EY 13337-03), and a Boswell Professorship (R.A.A). We acknowledge and thank Drs. John Pezaris and Maneesh Sahani for recording the LIP data which we used to test the algorithm. The authors declare that they have no competing financial interest.

FIGURE CAPTIONS:

Figure 1. Implementation of the algorithm for LFP recorded from area LIP. A. The time-frequency spectrum of the first principal component of the LFP recorded from a single site showing the underlying events at times $t_i$ with corresponding labels $a_i$. Also shown are windows $L_i$ used to compute feature vectors for each of these events. The windows are chosen to precede the events if the aim is to predict the events before they occur. B-E: Average spectra and the corresponding 2d cepstra computed for trials to the preferred direction (B,C) and trials to the anti-preferred direction (D,E). The vertical line in B,D denotes the beginning of the memory period. Note the enhancement of gamma band (25-90Hz) activity in the memory period for the preferred direction. The 2d cepstra also show differences between the two cases. In addition, they provide a compressed representation. F-G Detection of saccades to the preferred direction. F. First principal component of the LFP recorded at a single site along with a saccade to the preferred direction (P). G. Negative log-likelihood (risk) as a function of time for predicting the saccade P. Maximizing the log-likelihood is equivalent the minimizing the negative log-likelihood as discussed in the text. The local minimum at $b$ (for which the log-likelihood is greater than the threshold $\mu_P$) corresponds to a saccade at $S_P$ to the preferred direction. The feature vectors are computed with an offset $t_{Offset}$ relative to the saccades that are being predicted. The difference in time between $S_P$ and P is the error in the timing prediction. The prediction is accepted as correct if this difference is within a specified tolerance and provided it occurs separated from the previous prediction by a duration T.

Figure 2. Detection and prediction without externally supplied timing information. Average fraction of correct predictions for trials to the preferred direction (TP); expected number of predictions for a random detector (NP), and the attempt frequency (AF). The results are shown pooled across all recordings, and separately for the two monkeys. Note that the performance greatly exceeds that expected from a random detector. Note also that the performance of the LFP based predictions is comparable to that of the spikes, and that the performance of the LFP based predictions in recordings is not significantly affected by absence of single units. Error bars were estimated using jackknife over recordings. Values of TP and NP should be read on the left y-axis, while that of AF should be read on right y-axis.

Figure 3. Predictions for all recordings. Sorted values of the fraction of correct predictions (TP) for all recordings: 86 LFP recordings from sites without single units, 27 LFP recordings from sites with single units, and 30 single unit recordings. A. LFP recordings without single units, B. LFP recordings with single units, C. Single unit based predictions. Note the similarity in the distributions. The advantage of the LFP based predictions is that the number of recordings for which TP exceeds any specified value is higher for the LFP recordings than for the single unit recordings.

Figure 4. Histogram of timing differences. Timing differences between the predicted times of occurrence of events to the preferred direction and their true time. A. LFP, B. Single units. Both distributions are quite narrow with widths of roughly 70ms (LFP) and 85 ms (spikes).

Figure 5. Prediction of event type when timing information is supplied. Average percentage of correct predictions for saccades to the preferred direction (P); anti-preferred direction (A); all other directions (R). The results are shown pooled across all recordings, and separately for the two monkeys. Note that the performance of the LFP based predictions is comparable to that of the spikes. Note also that the performance of the LFP based predictions is not significantly affected by whether single units were present in the recordings. Error bars were estimated by a jackknife across recordings. The arrow at the bottom left of each plot indicates the noise level.

Figure 1

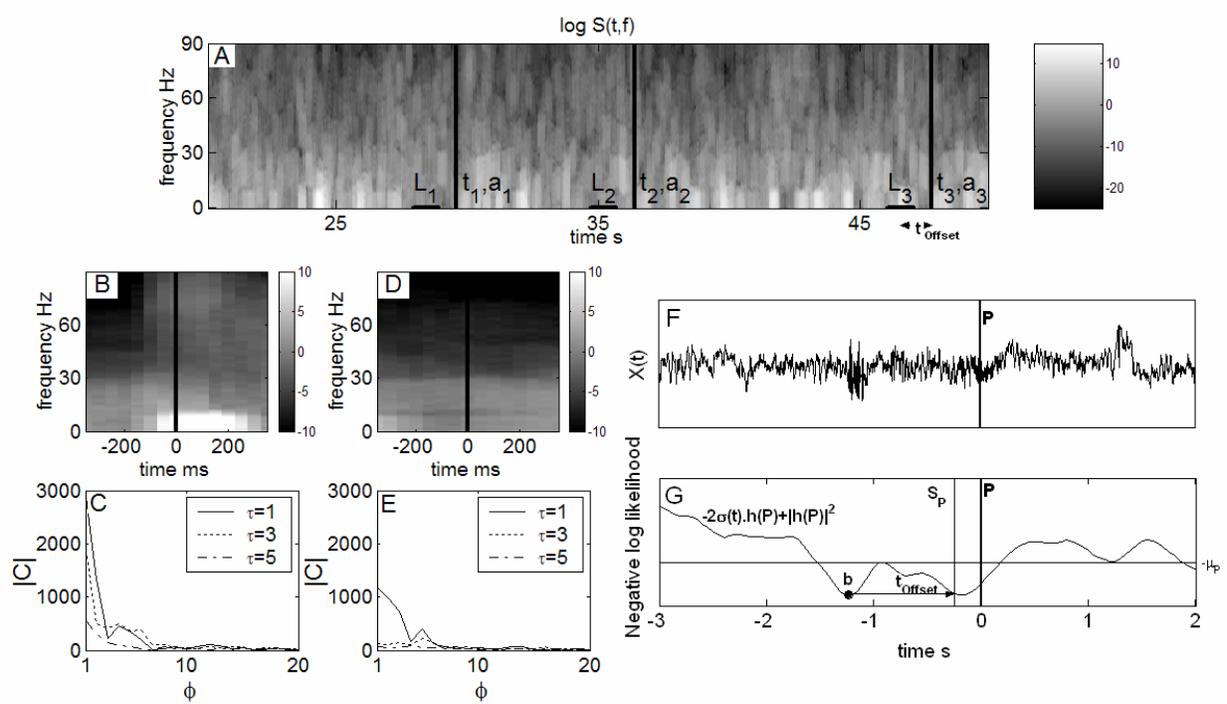

Figure 2

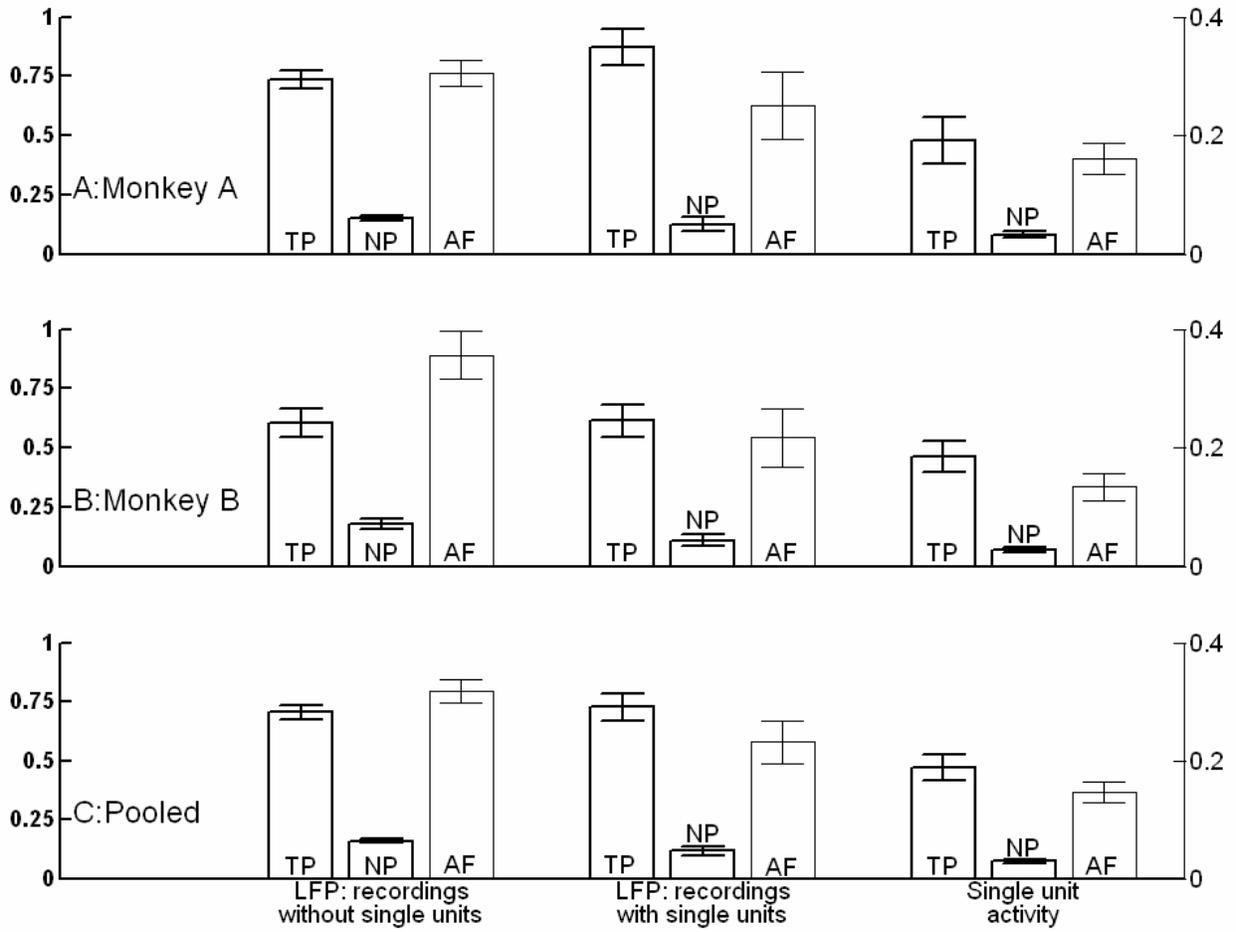

Figure 3

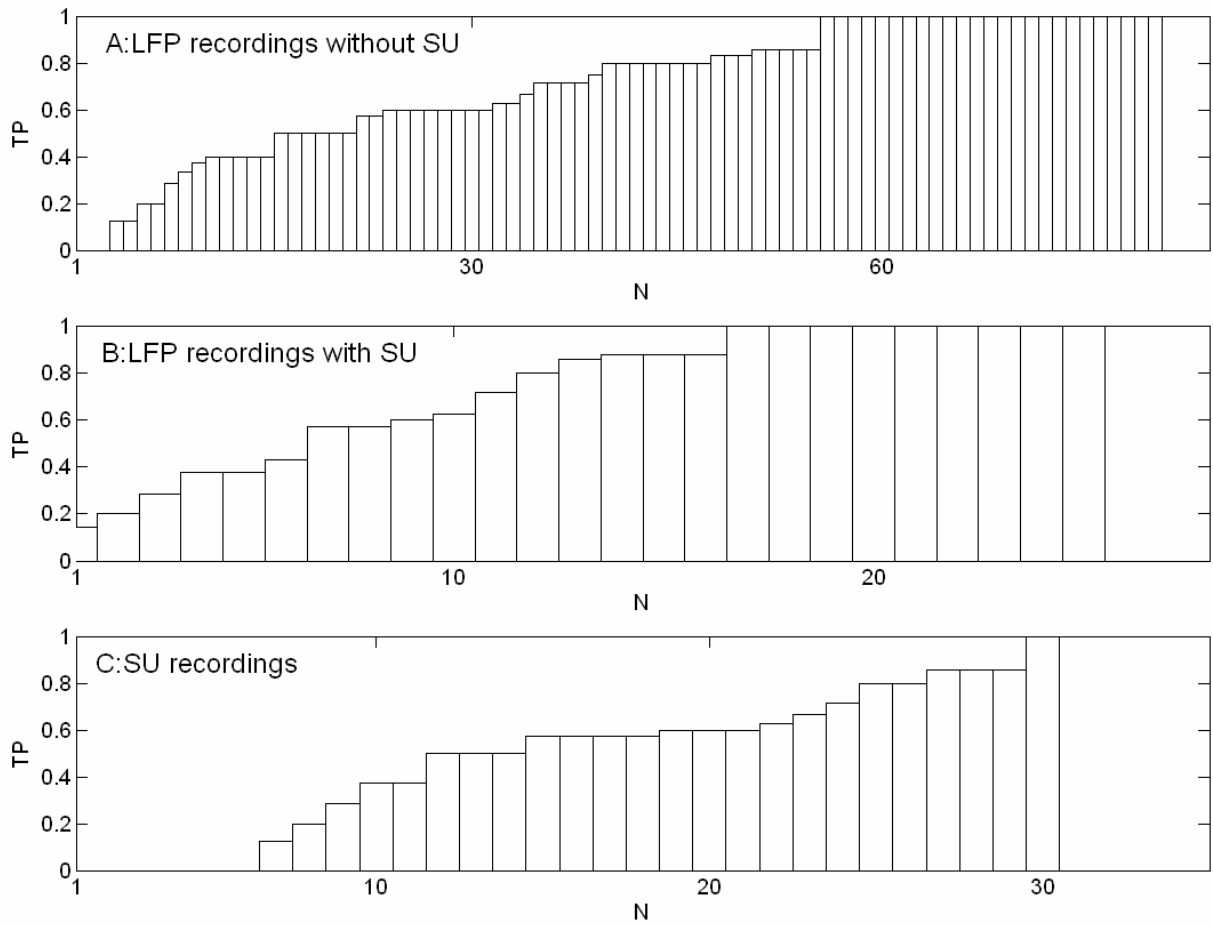

Figure 4

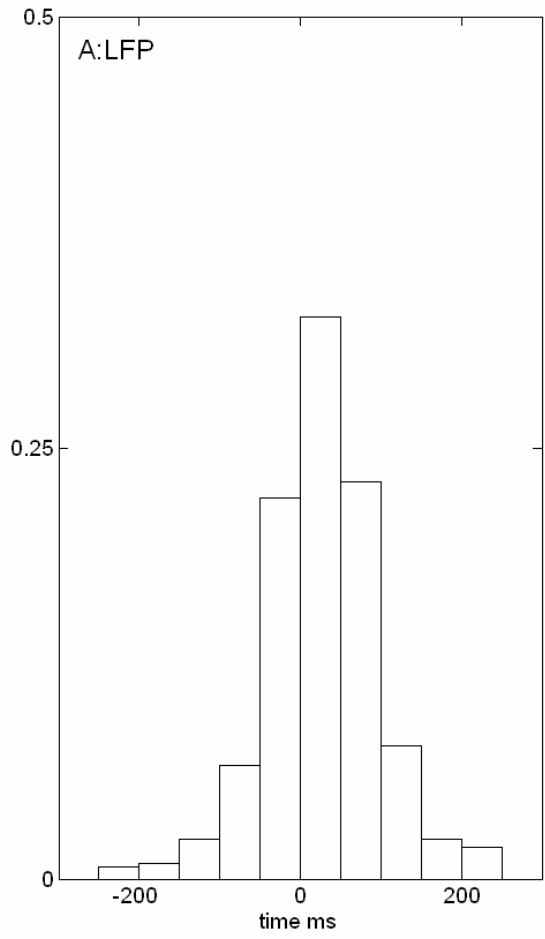
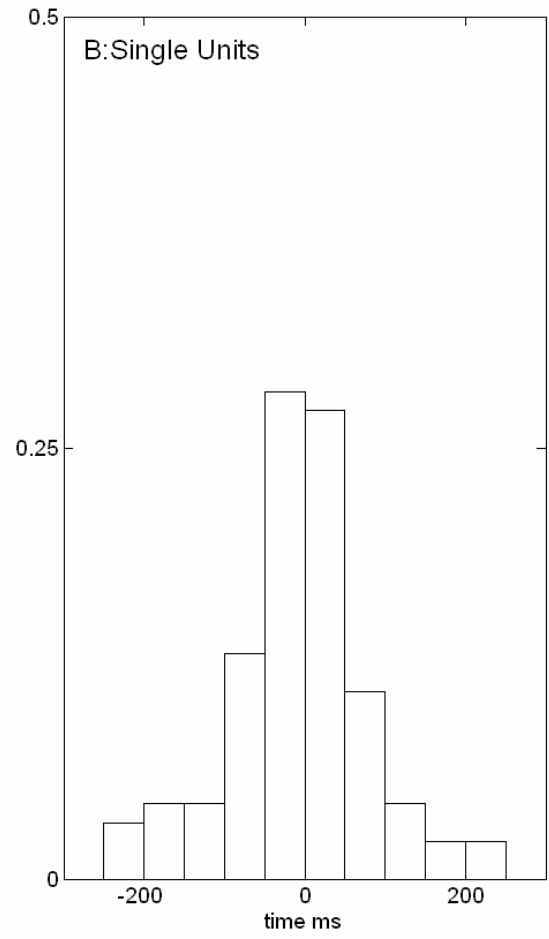

Figure 5

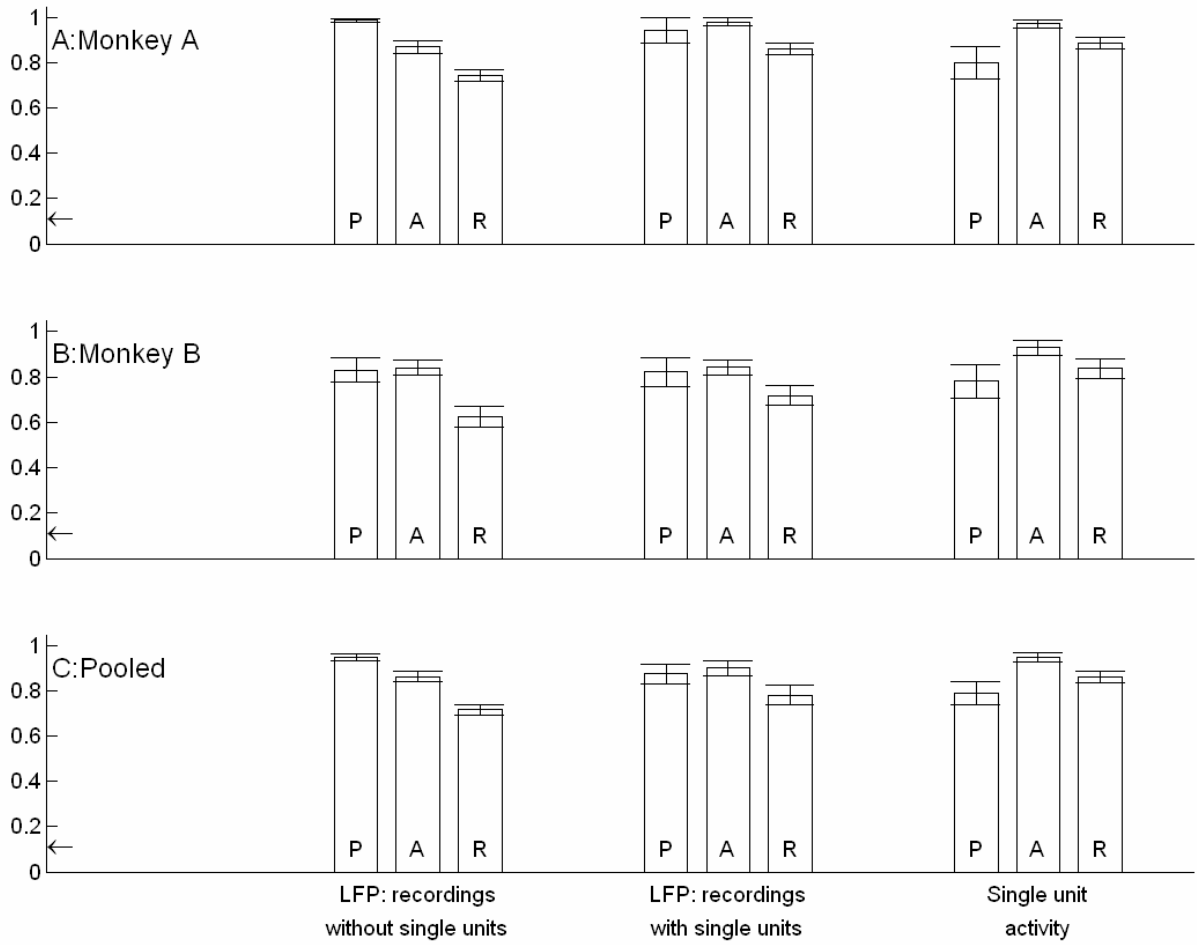